\documentclass[aip,apl,reprint,superscriptaddress
]{revtex4-1}
\usepackage{amsmath,amsfonts,amssymb}
\usepackage{xcolor}
\usepackage{epstopdf}
\usepackage{graphicx}
\renewcommand{\epsilon}{\varepsilon}
\renewcommand{\vec}{\bf}

\draft 

\begin{document}



\title{Interface band gap narrowing behind open circuit voltage losses in Cu$_2$ZnSnS$_4$ solar cells}






\author{Andrea Crovetto}
\email[]{Electronic mail: ancro@nanotech.dtu.dk}
\affiliation{DTU Nanotech, Technical University of Denmark, DK-2800 Kgs. Lyngby, Denmark}

\author{Mattias Palsgaard}
\email[]{Electronic mail: mattias.palsgaard@quantumwise.com}
\affiliation{DTU Nanotech, Technical University of Denmark, DK-2800 Kgs. Lyngby, Denmark}
\affiliation{QuantumWise A/S, DK-2100 Copenhagen, Denmark}

\author{Tue Gunst}
\affiliation{DTU Nanotech, Technical University of Denmark, DK-2800 Kgs. Lyngby, Denmark}

\author{Troels Markussen}
\affiliation{QuantumWise A/S, DK-2100 Copenhagen, Denmark}

\author{Kurt Stokbro}
\affiliation{QuantumWise A/S, DK-2100 Copenhagen, Denmark}
\author{Mads Brandbyge}
\affiliation{DTU Nanotech, Technical University of Denmark, DK-2800 Kgs. Lyngby, Denmark}

\author{Ole Hansen}
\affiliation{DTU Nanotech, Technical University of Denmark, DK-2800 Kgs. Lyngby, Denmark}
\affiliation{V-SUSTAIN, Villum Center for the Science of Sustainable Fuels and Chemicals, Technical University of Denmark, DK-2800 Kgs.~Lyngby, Denmark}

%

\begin{abstract}
We present evidence that band gap narrowing at the heterointerface may be a major cause of the large open circuit voltage deficit of Cu$_2$ZnSnS$_4$/CdS solar cells. Band gap narrowing is caused by surface states that extend the Cu$_2$ZnSnS$_4$ valence band into the forbidden gap. Those surface states are consistently found in Cu$_2$ZnSnS$_4$, but not in Cu$_2$ZnSnSe$_4$, by first-principles calculations. They do not simply arise from defects at surfaces but are an intrinsic feature of Cu$_2$ZnSnS$_4$ surfaces. By including those states in a device model, the outcome of previously published temperature-dependent open circuit voltage measurements on Cu$_2$ZnSnS$_4$ solar cells can be reproduced quantitatively without necessarily assuming a cliff-like conduction band offset with the CdS buffer layer. Our first-principles calculations indicate that Zn-based alternative buffer layers \textcolor{black}{are advantageous} due to the ability of Zn to passivate those surface states. Focusing future research on Zn-based buffers is expected to significantly improve the open circuit voltage and efficiency of pure-sulfide  Cu$_2$ZnSnS$_4$ solar cells.

\end{abstract}

\pacs{}

\maketitle 


Even though Cu$_2$ZnSnS$_4$ (CZTS) solar cells could be a sustainable solution to the increasing global energy demand, they are still plagued by a low open circuit voltage compared to their Shockley-Queisser limit, which has so far prevented them from reaching a sufficiently high efficiency for commercialization.\cite{Polman2016} In this work we show evidence of an interface mechanism limiting the open circuit voltage, and we demonstrate that this mechanism can be overcome by choosing a particular class of materials as interface partners of CZTS.

Knowledge of the "recombination energy deficit" $\Delta\phi$, i.e., the difference between the band gap of the absorber and the activation energy ($\phi$) of the main recombination path, can help identify where the limiting mechanism is located. $\Delta\phi$ can be estimated by a temperature-dependent open circuit voltage measurement.\cite{Redinger2013}
If $\Delta\phi > 0$, recombining electrons and holes are separated by an energy distance that is smaller than the absorber band gap. Here there is a significant difference between CZTS and its selenide equivalent Cu$_2$ZnSnSe$_4$ (CZTSe). State-of-the-art CZTSe solar cells are limited by bulk recombination because measured $\Delta\phi$ values correspond roughly to the depth of the bulk tail states
of CZTSe, from which carriers recombine.\cite{Redinger2013} Conversely, in state-of-the-art CZTS solar cells $\Delta\phi$ is around 0.4~eV, \cite{Wang2010,Tajima2015,Ericson2014} even though the depth of the CZTS bulk tail states is only 0.1-0.2~eV lower than the band gap.\cite{Shin2013,Platzer-Bjorkman2015} Such a mismatch implies that the energy distance between recombining electrons and holes is further reduced somewhere in the solar cell.
A popular hypothesis is that the interface between CZTS and its usual heterojunction partner CdS (or "buffer layer") features a cliff-like conduction band offset (CBO). In such a scenario, the energy distance between recombining electrons on the CdS side and holes on the CZTS side is reduced by an amount equal to the CBO. However, even though many reports of a cliff-like CBO exist for devices with efficiency below 5\%, \textcolor{black}{all} band alignment measurements on CZTS/CdS solar cells with efficiency above 7\% yielded a spike-like or \textcolor{black}{nearly} flat CBO\cite{Haight2011,Tajima2013,Kato2013,Terada2015} (Fig.~S1). Therefore, we conclude that \textcolor{black}{a large cliff-like CBO may exist in} some lower-performance CZTS/CdS solar cells but not in the best reported CZTS/CdS solar cells.

\begin{figure*}[t!]
\centering%
\includegraphics[width=2\columnwidth]{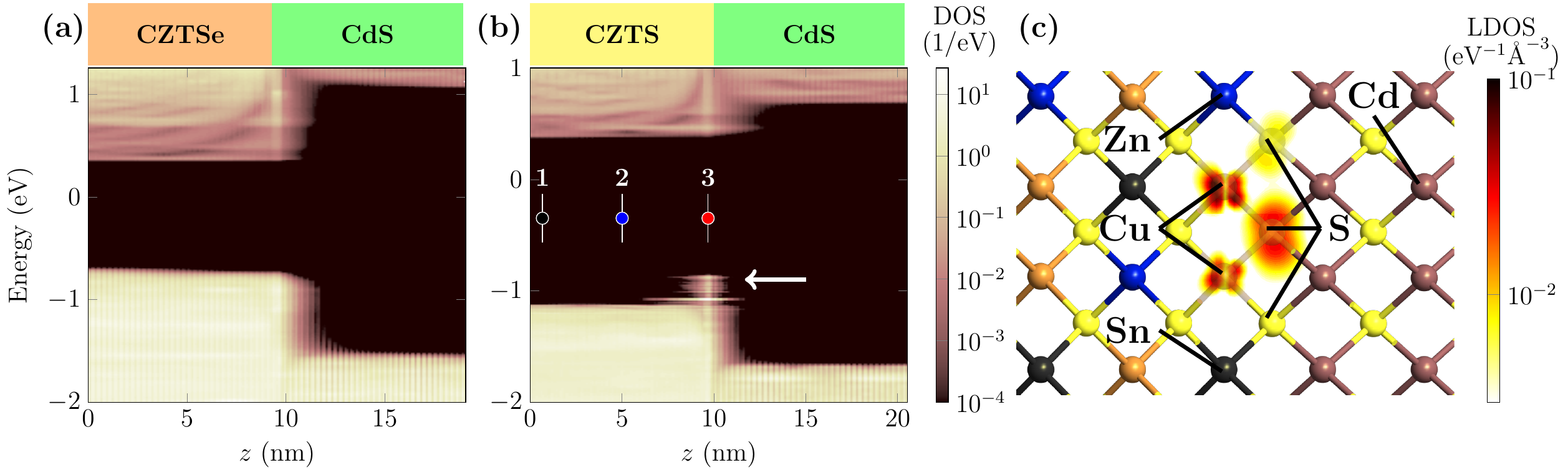}
\caption{Local density of states of the CZTSe/CdS interface (a) and of the CZTS/CdS interface (b) resolved along the direction perpendicular to the interface plane.
(c) Spatially-resolved DOS of the localized states at the CZTS/CdS interface. The DOS at a single energy, marked by the arrow in (b), is plotted in the figure.
}
\label{fig:spatial_dos}
\end{figure*}

\begin{figure}[t!]
\centering%
\includegraphics[width=1\columnwidth]{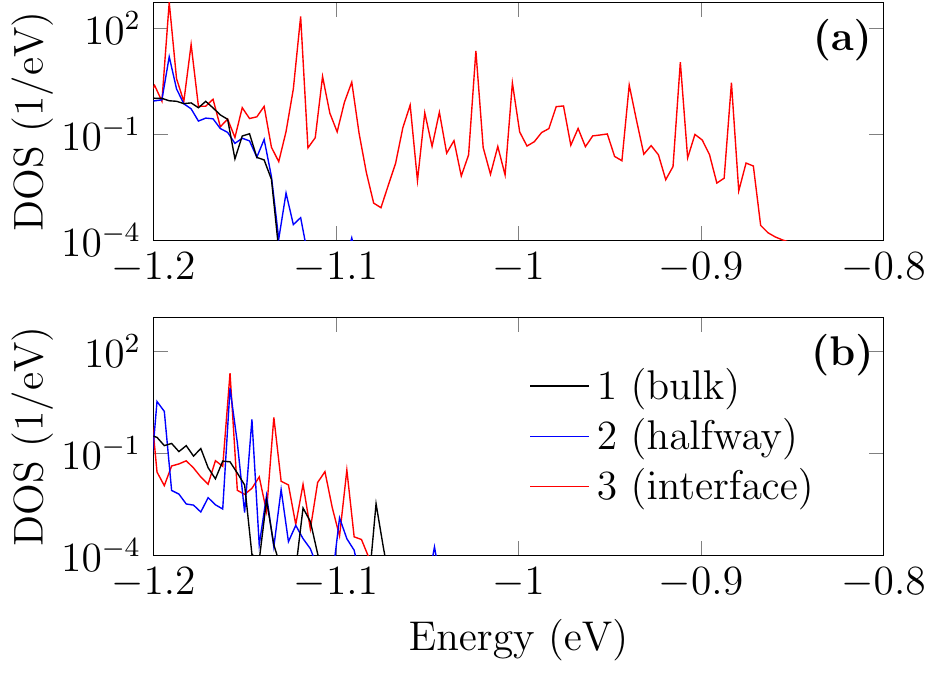}
\caption{Local density of states close to the VBM at the three positions indicated in Fig.~\ref{fig:spatial_dos}(b) for the case of: (a) the CZTS/CdS interface, and (b) the CZTS/ZnS interface.}
\label{fig:energy_dos}
\end{figure}


In an attempt to identify \textcolor{black}{another mechanism that may contribute to the large $\Delta\phi$ value}, we performed first-principles electronic structure calculations on the CZTS(100)/CdS(100) and CZTSe(100)/CdS(100) interfaces. The calculations were based on a density functional theory-nonequilibrium Green's function approach (DFT-NEGF) within the generalized gradient approximation (GGA-PBE) as implemented in the Atomistix ToolKit,~\cite{ATK} similarly to a previous publication.\cite{Palsgaard2016}
A semi-empirical Hubbard energy term was added to the GGA-PBE exchange-correlation potential to correct for self-interaction of localized d-orbitals and yield accurate band gaps (DFT+U approach).
All calculations were performed with a double-zeta-polarized basis set based on linear combination of atomic orbitals (LCAO).
Atomic positions of the CZTS(e) were relaxed keeping the experimental lattice parameters. For CdS we relaxed the atomic positions until all forces were below 0.02 eV/$\text{\AA}$ in a cell strained to fit that of CZTS(e) in the directions parallel to the interface. The lattice parameter perpendicular to the interface was relaxed until the stress was below 0.005 eV/\AA$^3$.
The choice of (100)/(100) interface orientation can be justified based on transmission electron microscopy results, which consistently show a (100)-oriented CZTS/CdS epitaxial interface.\cite{Tajima2014,Liu2016a} This also justifies modeling the CZTS/CdS interface as epitaxial in our calculation.
For the bulk calculations of CZTS (CdS) we used a $5 \times 5 \times 3$ ($5 \times 5 \times 5$) Monkhorst-Pack $\vec{k}$-point grid.
For the interface calculations (density of states calculations) we used $5 \times 3$ ($21 \times 21$) transverse $\vec{k}$-points.
As explained in a previous publication,~\cite{Palsgaard2016} a forward voltage bias was applied across the supercell to remove residual slopes of the local potential near the electrodes. For an appropriate magnitude of the applied voltage, flat-band conditions are achieved. To justify this approach, we emphasize that: i) the electrostatic potential drop at the junction (band bending) occurs over a much larger length scale than the supercell length,\cite{Shin2013,Sun2016,Tajima2015} thus to a first order approximation the bands can be assumed to be flat within the calculated region; ii)
the optimal working point of the solar cell device is indeed close to flat-band conditions (forward bias).

Fig.~\ref{fig:spatial_dos}(a,b) shows the calculated density of states (DOS) close to the band edges in the interface region for the two materials pairs.
The resulting CBOs are +0.2~eV for the CZTS/CdS interface and +0.6~eV for the CZTSe/CdS interface.
Besides the differences in band alignment, we note that localized interface states are present at the CZTS/CdS interface but are absent from the CZTSe/CdS interface.

The existence of the localized states results essentially in an extension of the valence band up to 0.2-0.3 eV above the original valence band maximum (VBM) of CZTS (Fig.~\ref{fig:energy_dos}(a)). By repeating the calculation using different computational techniques and modeling assumptions, we have verified that
the presence of the localized states (and their absence at the CZTSe/CdS interface) is not an artifact of the calculation. This is shown in the supplementary material.
In a separate calculation on a S-terminated CZTS/vacuum interface ("surface calculation"), localized states above the VBM of CZTS were also observed (Figs. S6, S7). This suggests that the states are due to dangling bonds at the CZTS surface that are not satisfactorily passivated by a CdS buffer layer.
%
In fact, the states are highly localized on Cu sites in the first cationic layer of CZTS and on their neighboring S atoms in the interface anionic layer (Fig.~\ref{fig:spatial_dos}(c)). Since the valence band of CZTS originates from Cu and S states, \cite{Palsgaard2016,Persson2010} this explains why those localized states affect the valence band but not the conduction band.
Interestingly, there exists some experimental evidence of the presence of electrically active surface states in CZTS and of their absence from CZTSe, as predicted by our calculation. A surface photovoltage measurement by scanning tunneling microscopy \cite{Du2011} revealed that, in CZTSe, the photocurrent scaled linearly with optical excitation intensity, whereas in CZTS the photocurrent saturated quickly. The authors concluded that this was due to the predominance of surface states in the CZTS response but not in the CZTSe response. In another study, a work function measurement on CZTSSe surfaces with inhomogeneous S/(S+Se) content revealed that the Fermi level position in areas of higher S content did not match the theoretical expectation based on the band edge positions of ideal bulk materials.\cite{Salvador2014}

The identification of localized states at the CZTS/CdS interface can help explain
why state-of-the-art CZTS/CdS solar cells still have a large $\Delta\phi$ even in the case of an optimal band alignment with CdS. To demonstrate this quantitatively, we incorporate the first-principles calculation results into a model for device-level simulation. Simulation of a CZTS/CdS/ZnO solar cell was carried out with the finite element method as implemented in the software SCAPS.\cite{Burgelman2000} Device parameters are listed in Table~S1. Besides the inclusion of the localized stats, the device model has two important features. The first is the small spike-like band alignment between CZTS and CdS, consistent with state-of-the-art CZTS/CdS solar cells (Fig.~S1). The second is the distinction between an optical band gap of 1.5~eV and a transport band gap of 1.35~eV. This simulation approach has been suggested before\cite{Frisk2016} to model the mismatch between the band gap of the extended states and the band gap from which bulk recombination occurs (which includes the tail states due to bulk fluctuations in the CZTS band edges). 0.15~eV is a typical depth for the tail states of high-quality CZTS.\cite{Shin2013,Platzer-Bjorkman2015}
\\
The interface states are included in the device model as follows. According to the first-principle calculations, the states have a similar DOS to the valence band of bulk CZTS up to 0.2-0.3~eV above the original VBM (Fig.~\ref{fig:energy_dos}(a)) and are only present in an interface region that extends less than 5~nm into CZTS (Fig.~\ref{fig:spatial_dos}(b)). Therefore, the interface states are modeled as a 0.2~eV upward shift in the valence band over a 5 nm region at the interface, rather than a valence band tail or a single defect level within the gap. This is equivalent to narrowing the interface band gap on the CZTS side of the junction, which means that the energy barrier for recombination is reduced (by 0.2~eV) at the interface, much like the case of a cliff-like CBO with CdS.
The other materials parameters of this interface region are kept identical to a baseline CZTS device without interface states that we simulated for comparison.

\begin{figure}[t!]
\centering%
\includegraphics[width=1\columnwidth]{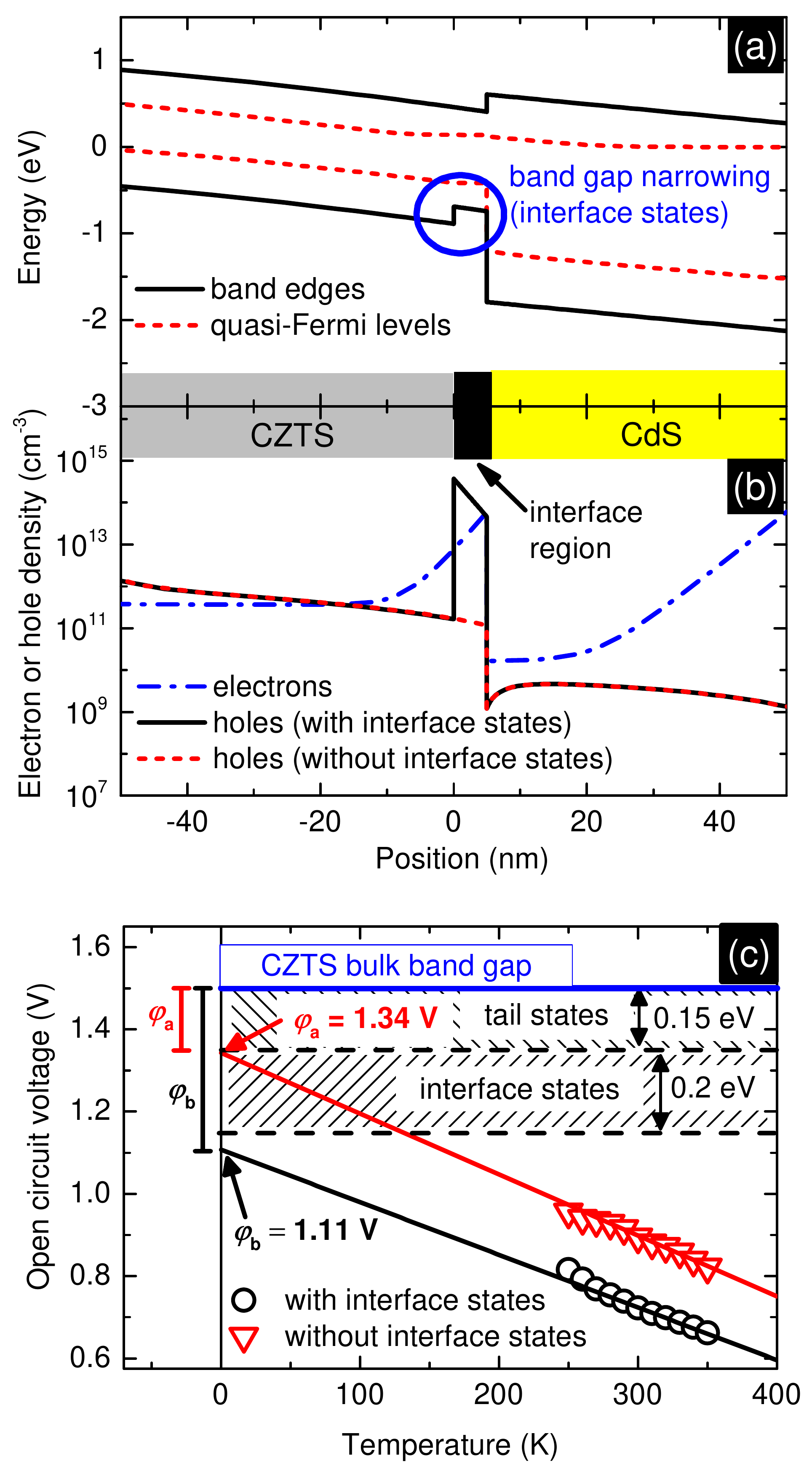}
\caption{Simulated properties of the interface region of a CZTS/CdS solar cell under AM1.5 illumination, with the inclusion of localized states at the interface.
(a) Band diagram and corresponding quasi-Fermi levels for electrons and holes.
(b) Electron and hole density in the same region.
The hole density increases significantly when interface states are added, whereas the electron density is the same in both scenarios.
(c) Simulated open circuit voltage of CZTS/CdS solar cells as a function of temperature. The linearly extrapolated 0~K intercept of the data yields the activation energy of the dominant recombination path.}
\label{fig:device_simulation}
\end{figure}


The near-interface band diagram of the simulated device is shown in Fig.~\ref{fig:device_simulation}(a). In a device with interface states, the interface hole density increases by three orders of magnitude compared to the baseline device, up to a range that is comparable to the electron density (Fig.~\ref{fig:device_simulation}(b)). This is a crucial effect that implies a higher Shockley-Read-Hall interface recombination rate, because the latter increases with increasing electron and hole densities, and is maximized when the two densities equal each other.\cite{Sah1957}
To verify that this device model is compatible with real CZTS/CdS solar cells, we use our device model to simulate a temperature-dependent open circuit voltage measurement from which $\phi$ is usually extracted experimentally (Fig.~\ref{fig:device_simulation}(c)). 
\textcolor{black}{Open circuit voltages are obtained by simulating the current-voltage (JV) curve of the solar cell under AM1.5 illumination at different simulated temperatures (Fig.~S9).}
For the baseline case without interface states, extrapolation of the open circuit voltage to 0~K yields 1.34~V. This matches the value of the bulk transport band gap (1.35~eV) defined in our model, and it means that such a baseline device is not limited by interface recombination.
In the solar cell with interface states, the open circuit voltage extrapolates to 1.11~V, which matches the value of the transport gap minus 0.2~eV narrowing at the interface as defined for the interface region (1.15~eV). The corresponding recombination energy deficit is 0.39~eV, which fits very well the recombination energy deficits of 0.3~eV, 0.4~eV, and 0.4~eV found experimentally in the highest-efficiency CZTS/CdS solar cells\cite{Tajima2015,Wang2010,Ericson2014}.
\textcolor{black}{As long as the simulated device is dominated by interface recombination, the simulated value of the recombination energy deficit is robust with respect to changes in various device parameters, including defect characteristics.}

Such findings demonstrate that narrowing of the interface band gap through an upward shift of the CZTS valence band can explain existing temperature-dependent open circuit voltage measurements just as well as a cliff-like CBO does. They also imply that optimal passivation of the interface states can result in a considerable enhancement of the open circuit voltage. This interesting prospect may be practically realized by replacing CdS with an appropriate passivation material. The question is which material would work. Here we limit our analysis to a (100)/(100) interface with a metal chalcogenide (MX, where M is the metal and X is the chalcogen) with cubic or tetragonal structure. The interfacial cationic layer of such a material breaks the bulk crystal structure of CZTS by introducing a layer of $2\mathrm{M}_{\mathrm{Cu}}+\mathrm{M}_{\mathrm{Zn}}+\mathrm{M}_{\mathrm{Sn}}$ point defects.
Therefore, one strategy could be to search for a metal whose related defect complex does not form states within the band gap of CZTS. If one chooses $\mathrm{M}=\mathrm{Zn}$, the defect complex reduces to just $2\mathrm{Zn}_{\mathrm{Cu}}+\mathrm{Zn}_{\mathrm{Sn}}$. The effect of this particular defect complex on the electronic properties of CZTS has been investigated before.\cite{Chen2013} The result was that the $2\mathrm{Zn}_{\mathrm{Cu}}+\mathrm{Zn}_{\mathrm{Sn}}$ complex does not narrow the band gap of CZTS. Therefore, one may expect a Zn chalcogenide material (such as ZnS) to remove the interface states.

To test this hypothesis, we repeated our CZTS interface calculation replacing CdS with ZnS (the computational methods involving ZnS are the same as for CdS). Strikingly, Fig.~\ref{fig:energy_dos}(b) shows that no interface states are present anymore, within the resolution of the calculation.
This indicates that the ideal situation of a CZTS interface without band gap narrowing could be achieved by replacement of CdS by a Zn chalcogenide, with a corresponding shift of the dominant recombination path from the interface to the bulk. Our results provide a clear explanation of why open circuit voltage improvement due to interface modification has so far been achieved experimentally by Zn-based alternative buffers Zn$_{1-x}$Sn$_{x}$O$_x$,\cite{Platzer-Bjorkman2015} (Zn,Cd)S,\cite{Sun2016} and another unspecified Zn-based buffer.\cite{Sakai2011}
Zn$_{1-x}$Sn$_{x}$O$_x$ has also been the only material able to reduce $\Delta\phi$ from the typical 0.3-0.4~eV down to 0.16~eV,\cite{Platzer-Bjorkman2015} which corresponds roughly to the depth of CZTS bulk tail states.

To conclude, we have shown that the interface band gap of CZTS/CdS solar cells is narrowed by localized states that shift the valence band maximum of CZTS to a higher energy. The same effect does not occur at the CZTSe/CdS interface.
This phenomenon can explain why interface recombination is always observed to dominate in CZTS solar cells but not in CZTSe solar cells -- a fact that has so far been attributed to differences in the conduction band alignment of the two materials with the CdS buffer layer. Zn-based chalcogenides can effectively passivate CZTS surfaces by removing the localized states. This can explain why Zn-based alternative buffer layers have \textcolor{black}{so far outperformed other buffer layer materials}, thus giving a clear recipe for future improvement.
\vspace{0.5cm}

See supplementary material for a review of the band alignment between CZTS and CdS; confirmation of the interface band gap narrowing phenomenon by alternative methods; atomic structures used to calculate electronic properties; spatial density of states of a calculated CZTS/ZnS interface; a table with all parameters used in device simulation; \textcolor{black}{and simulated current-voltage curves}.

\vspace{0.5cm}
This work was supported by the Danish Council for Strategic Research, VILLUM Fonden (grant 9455) and the Innovation Fund Denmark (File No. 5016-00102).


%

\end{document}